\documentclass[prb,aps,twocolumn,amsmath,amssymb,floatfix,
superscriptaddress]{revtex4}

\usepackage[dvips]{graphics}
\usepackage{color}
\definecolor{dred}{rgb}{0.75,0,0}

\begin{document}

\title{Unconventional localization phenomena in a spatially non-uniform 
disordered material}

\author{Madhuparna Banerjee}

\affiliation{Department of Physics, Budge Budge Institute of Technology, 
Nischintapur, Budge Budge, Kolkata-700 137, India}

\author{Baisakhi Mal}

\affiliation{Department of Physics, Heritage Institute of Technology,
Chowbaga Road, Anandapur, Kolkata-700 107, India}

\author{Santanu K. Maiti}

\email{santanu.maiti@isical.ac.in}

\affiliation{Physics and Applied Mathematics Unit, Indian Statistical
Institute, 203 Barrackpore Trunk Road, Kolkata-700 108, India}

\begin{abstract}

A completely opposite behavior of electronic localization is revealed in a 
spatially non-uniform disordered material compared to the traditional 
spatially uniform disordered one. This fact is substantiated by considering 
an order-disorder separated (ODS) nanotube and studying the response of 
non-interacting electrons in presence of magnetic flux. We 
critically examine the behavior of flux induced energy spectra and circular 
current for different band fillings, and it is observed that maximum current 
amplitude (MCA) gradually decreases with disorder strength for weak disorder 
regime, while surprisingly it (MCA) increases in the limit of strong disorder 
suppressing the effect of disorder, resulting higher conductivity. This is 
further confirmed by investigating Drude weight and exactly same anomalous 
feature is noticed. It clearly gives a hint that localization-to-delocalization 
transition (LTD) is expected upon the variation of disorder strength which 
is justified by analyzing the nature of different eigenstates. Our analysis 
may give some significant inputs in analyzing conducting properties of 
different doped materials.

\end{abstract}

\maketitle

\section{Introduction}

The phenomenon of electronic localization is still alive with great interest
in the discipline of condensed matter physics since its prediction in 1958 
by P. W. Anderson~\cite{ander}. It is well known that for an {\em infinite} 
one-dimensional (1D) random (uncorrelated) disordered lattice all states are 
localized irrespective of disorder strength $W$~\cite{ander,lee}. As for this 
system the critical disorder strength is zero one never expects any kind of 
localization-to-delocalization transition upon the variation of $W$, which
circumvents the appearance of mobility edge (ME)~\cite{choi,san1,skm1} in energy 
band spectrum that 
separates a conducting zone from the localized one. The existence of mobility 
edge always draws significant impact particularly in the aspect of its 
applications in designing possible electronic devices like switching action,
current transfer processes, and to name a few. To expect mobility edge(s) 
one needs to go beyond 1D `random' disordered model. There is a specific
class of 1D materials known as Aubry-Andr\'{e} or Harper (AAH) model where
mobility edge can be observed under certain conditions~\cite{aub}. The systems 
belong to this class are no longer random disordered one, rather they are 
deterministic since site energies and/or hopping integrals follow a specific
functional relation. This AAH model is a classic example where extended or 
localized or mixture of both energy eigenstates are obtained depending of 
the suitable parameter range. A wealth of literature knowledge has already
been developed in such models with considerable theoretical and experimental 
works~\cite{mou,das1,das2,exp1,exp2,exp3,exp4}.

Now if we stick to the `random' disordered model and want to observe the 
phenomenon of LTD transition we need to switch to the higher-dimensional one. 
In a pioneering work Anderson has shown~\cite{ander} that for the 
three-dimensional (3D) random disordered system mobility edge can be observed, 
but the important restriction is that the disorder strength should be weak. 
So the question naturally comes {\em can we think about a system with 
uncorrelated site energies that can exhibit mobility edge phenomenon and 
persist even at higher disorder limit.} The answer is yes, and, it can be 
implemented by considering a {\em spatially non-uniform uncorrelated disordered 
system} unlike the conventional disordered one where uncorrelated site energies 
are distributed 
uniformly throughout the material. This type of materials, for instance 
spatially doped semi-conducting materials, doped nanotubes and nanowires,
and many more, has nowadays been accessible quite easily, particularly 
\begin{figure}[ht]
{\centering \resizebox*{5cm}{7cm}{\includegraphics{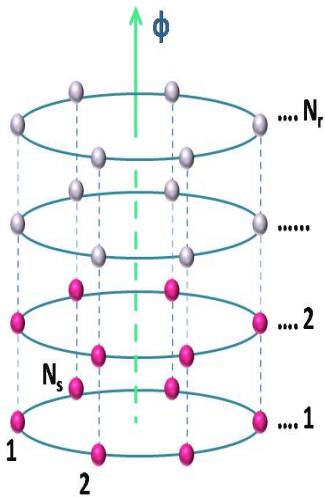}}\par}
\caption{(Color online). Schematic diagram of a single-wall order-disorder
separated nanotube subjected to a magnetic flux $\phi$. The impurities are 
distributed only in the filled red circles keeping the other sites (white
circles) free from impurities. $N_r$ corresponds to the total number of 
rings where each of these rings contains $N_s$ atomic sites. In presence 
of $\phi$ a net circular current is established.}
\label{fig1}
\end{figure}
because of enormous progress in designing nanoscale systems. Several diverse 
characteristics are observed~\cite{div1,div2,div3,div4,div5,div6,div7}, as 
reported mostly by experimental works with too less theoretical attempts, as 
far as we know, in different kinds of doped semi-conducting materials which 
essentially trigger us to probe into it further. 

In the present work we do an in-depth analysis of magneto-transport properties 
of non-interacting electrons in an order-disorder separated single-wall 
nanotube (SWNT). Our focus of this work is essentially two-fold. First, we 
want to consider a simple geometrical structure that may capture the essential 
physics of different 
kinds of doped materials. In our case it is an ODS SWNT. Second, we intend to 
discuss explicitly the localization behavior by studying magnetic response of 
non-interacting electrons, a different way compared to the established results.
The detailed investigation of localization phenomena in a spatially non-uniform
disordered system by analyzing magnetic response has not been probed so far. 
In presence of magnetic flux $\phi$ a net circular current is established in the 
nanotube, a direct consequence of Aharonov-Bohm (AB) effect, and it does not 
vanish even after removing the flux. This is the well-known phenomenon of 
flux-induced persistent current in mesoscopic AB loops and was first predicted 
theoretically by B\"{u}ttiker 
and his group~\cite{pc1}. Soon after this prediction interest in this topic 
has rapidly grown up with significant theoretical and experimental works 
considering different kinds of single and multiple loop geometries~\cite{pc2,pc3,
pc4,pc5,pc6,pc7,pc8,pc8a,pc9,pc10,pc11,skm2,skm3,skm5}. Various physical 
properties like conducting nature of full system along with individual energy
eigenstates, magnetization behavior, etc., can be analyzed directly by 
studying persistent current. The role of disorder has always been an 
interesting issue as it is directly involved with electronic localization 
that affects significantly the current amplitude. The conventional notion
suggests that for a fixed band filling current amplitude gradually decreases
with disorder strength. Unlike this a spatially non-uniform disordered system 
exhibits anomalous behavior beyond a critical disorder strength where current
amplitude increases suppressing the effect of disorder, yielding higher 
electrical conductivity which is confirmed by studying Drude 
weight~\cite{kohn,bouz,skm4}. 
The non-uniform spatial distribution of impurities plays the central role of all
these atypical signatures which we justify by critically investigating the
nature of different energy eigenstates. For the weak coupling limit the 
system behaves like a traditional disordered system, while in the limit of 
strong disorder two regions (viz, ordered and disordered sectors) behave 
completely differently. As a result of this, mobility edge phenomenon is 
observed which yields LTD transition, and unlike uniformly distributed
3D random disordered system, it persists even for strong disorder regime. 

The work is organized as follows. In Sec. II we present the model and 
theoretical formulation for the calculation of persistent current at 
different band fillings, Drude weight and inverse participation ratio (IPR). 
The results are analyzed in Sec. III, and finally, we conclude in Sec. IV.

\section{Model and Theoretical Formulation}

Let us begin with Fig.~\ref{fig1} where a single-wall nanotube
subjected to a magnetic flux $\phi$ is shown. The SWNT consists of $N_r$ 
number of vertically attached co-axial rings where each ring contains 
$N_s$ lattice sites. In order to get an order-disorder separated nanotube
we introduce impurities in one portion which is lower half of the SWNT 
without disturbing the other half part (i.e., upper half), and, throughout 
the analysis we follow this prescription for ODS nanotube. For fully 
disordered nanotube we add impurities uniformly all along the nanotube,
and it becomes a conventional disordered one. The perfect nanotube is the
trivial one where impurities are no longer introduced and all sites are
identical. All these three types of nanotubes can be simulated by a general 
tight-binding (TB) Hamiltonian~\cite{pc2} which looks like
\begin{eqnarray}
H & = &  \sum_{i,j} \epsilon_{i,j} \textbf{c}^\dagger_{i,j}\, 
\textbf{c}_{i,j} + \sum_{\langle i,j \rangle} 
\left[t_y (\textbf{c}^\dagger_{i,j} \textbf{c}_{i,j+1} +
 \textbf{c}^\dagger_{i,j+1} \textbf{c}_{i,j})
 \right. 
\nonumber \\
&& \left. + \, t_x  (e^{i\theta} \, \textbf{c}^\dagger_{i,j}\,
\textbf{c}_{i+1,j} + e^{-i\theta} \textbf{c}^\dagger_{i+1,j} 
\textbf{c}_{i,j})\right]
\label{eq1}
\end{eqnarray}
where $(i,j)$, position of a lattice site, represents $i$th site in the 
$j$th ring of the SWNT, and, $i$ and $j$ run from $1$ to $N_s$ and $N_r$, 
respectively. $\epsilon_{i,j}$ describes on-site energy which we choose 
`randomly' from a `Box' distribution function of width $W$ (i.e., 
$\epsilon_{i,j}$ lies within the range $-W/2$ to $+W/2$) for impurity sites, 
and this parameter $W$ measures the strength of disorder. Whereas, for the
perfect atomic sites $\epsilon_{i,j}$'s are same and we set them to zero 
(viz, for this case $W=0$) without loss of generality. $t_x$ and $t_y$ are the 
intra-and inter-channel nearest-neighbor hopping (NNH) integrals, respectively,
and because of the magnetic flux $\phi$ a phase factor $\theta$ 
($= 2\pi \phi /N_s$) appears into the Hamiltonian when an electron hops 
along the ring. No such phase factor appears for the hopping from one
ring to the other. $\textbf{c}_{i,j}$ and $\textbf{c}^\dagger_{i,j}$ are
the usual Fermionic operators.

This is all about the TB Hamiltonian of the SWNT. Now to inspect its physical
properties first we need to find out allowed energy levels, and for a perfect
SWNT we do it completely analytically, while in presence of impurities the
eigenvalues are determined by diagonalizing the Hamiltonian matrix of the TB
Hamiltonian Eq.~\ref{eq1}. For a fully ordered SWNT the energy eigenvalues 
are expressed as,
\begin{eqnarray}
E_{j,i} &=& 2t_y \cos\left(\frac{j\pi}{N_r + 1}\right)  
+ 2 t_x \cos \left[\frac{2\pi}{N_s} \left(i + \frac{\phi}{\phi_0}\right) 
\right]
\label{eq2}
\end{eqnarray}
where $-N_s/2 \le i < N_s/2$ and $j = 1,2,\dots, N_r$. Once eigenvalues
are found out, the total energy $E_0(\phi)$ for fixed number of electrons 
$N_e$ at absolute zero temperature can be obtained by taking the sum of
lowest $N_e$ energy levels, and the persistent current is evaluated from
the relation~\cite{pc2} 
\begin{equation}
I = -c\frac{\partial E_0(\phi)}{\partial \phi}
\label{eq3}
\end{equation}
This is the simplest way (viz, the derivative method) of calculating persistent 
current. Some other approaches are also available~\cite{skm2} and one can 
equally utilize anyone of them, but those techniques are normally used for
calculating response at different branches of a full system, which is not 
required for our present analysis. 

We investigate conducting properties of SWNT by determining Drude weight
$D$ which is expressed as~\cite{kohn,bouz}
\begin{equation}
D = \left.\left(\frac{N}{4\pi^2}\right) 
\frac{\partial^2 E_0(\phi)}{\partial\phi^2} \right\vert_{\phi\rightarrow 0^+}
\label{eq4}  
\end{equation}
where $N$ ($=N_s\times N_r$) being the total number of lattice sites in the tube.
Finite $D$ suggests the conducting phase, while for the insulating phase
$D\rightarrow 0$. The idea of estimating conducting properties by studying
Drude weight was originally put forward by Kohn~\cite{kohn}, and latter many 
other groups have discussed about it in detail~\cite{sca1,sca2}. This is a 
very good prescription, specially for isolated system (i.e., the system without 
any external baths).

Finally, to characterize mobility edge phenomenon we calculate inverse 
participation ratio of different energy eigenstates, and for any particular 
eigenstate $|\psi_n\rangle$ (say) it becomes~\cite{das2}
\begin{equation}
\mbox{IPR}=\frac{\sum\limits_{i,j} |a_n^{i,j}|^4}{\left(\sum\limits_{i,j}
|a_n^{i,j}|^2\right)^2}
\label{eq5}
\end{equation}
where $a_n^{i,j}$'s are the coefficients. For the localized state IPR is 
finite and its maximum possible value can be unity, while for an extended 
state it (IPR) becomes too small and drops to zero in the asymptotic 
limit~\cite{das2}.

\section{Results and discussion}

In what follows we present our results. Throughout the analysis we set 
the intra- and inter-channel NNH integrals at $-1\,$eV and fix the site 
energy to zero for ordered atomic sites. For disordered regions site energies
are random and since they are random we compute the results averaging over
a large number of disordered configurations to authenticate all the 
characteristic features. As the system temperature does not have any 
significant impact on our present analysis, we fix it at absolute zero, and
also set $c=e=h=1$ for the entire discussion, for simplification.

\begin{figure}[ht]
{\centering \resizebox*{8cm}{13cm}{\includegraphics{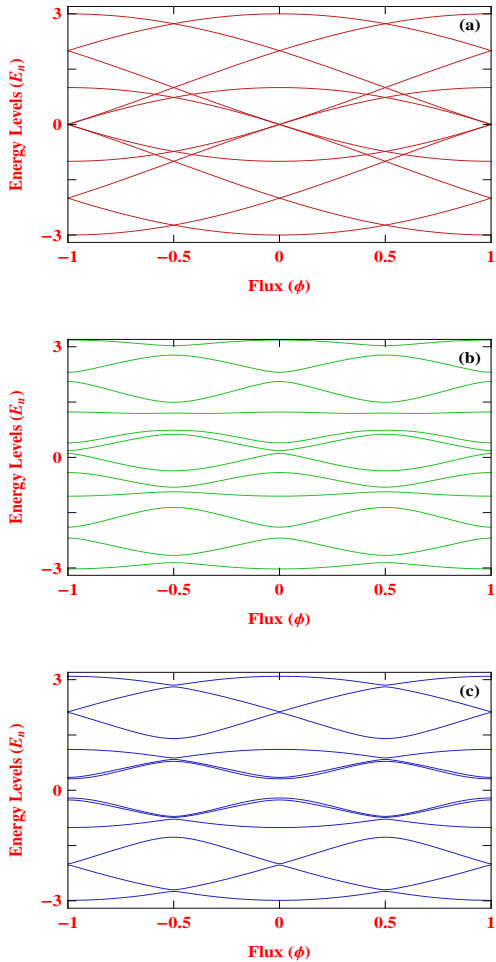}}\par}
\caption{(Color online). Variation of different energy levels as a function 
of magnetic flux $\phi$ for three different cases where (a), (b) and (c)
correspond to fully perfect, fully disordered and order-disorder separated
SWNTs, respectively, considering $N_s = 6$ and $N_r = 2$. For the disorder
sector we set $W = 2$.}
\label{fig2}
\end{figure}

Before addressing the central part i.e., unconventional behavior of circular 
current and the appearance of phase transition from conducting to insulating
zone upon the variation of impurity strength in an ODS SWNT, let us focus
on the energy band spectrum and current-flux characteristics for some 
typical electron fillings of different SWNTs to make the present 
communication a  self-contained one. In Fig.~\ref{fig2} we present
the variation of different energy levels with magnetic flux $\phi$
for three different SWNTs, fully ordered, fully disordered and ODS, 
considering $N_s=6$ and $N_r=2$.
From now on we refer energy levels as $E_n$ instead of using two indices $i,j$
(viz, $E_{i,j}$) to read it quite simply and with this notation no physics will
be altered anywhere. From the energy spectra it is observed that the fully 
ordered SWNT exhibits intersecting energy levels (Fig.~\ref{fig2}(a)) where 
the intersection takes 
place at integer and/or half-integer multiples of flux-quantum $\phi_0$, which 
may yield a sharp jump in current-flux characteristics at these typical fluxes 
though it eventually depends on the electron filling. The situation is somewhat
different if we introduce impurities. Both for fully disordered and ODS SWNTs
all such crossings of energy levels as noticed in fully ordered SWNT completely
disappear and energy levels vary continuously with flux $\phi$ 
(Figs.~\ref{fig2}(b) and (c)). At a first glance it is quite difficult to find
any difference between the spectra given in Figs.~\ref{fig2}(b) and (c), but
a careful inspection reveals that for a fully disordered SWNT slopes of the 
energy levels with flux $\phi$ are lesser than the ODS nanotube, and for some
energy levels it is too small, seems almost flat, for which the current will 
almost vanish. With increasing disorder strength $W$ flatness will be more 
prominent in both these two SWNTs, but for an ODS SWNT we always get some
energy levels which exhibit higher slopes which we confirm through our extensive
numerics. This definitely gives a hint that for a fully disordered nanotube 
current gets reduced with $W$, while for an ODS nanotube it may not be the case
and from our forthcoming analysis the nature will be clearly understood.

\begin{figure}[ht]
{\centering \resizebox*{8cm}{13cm}{\includegraphics{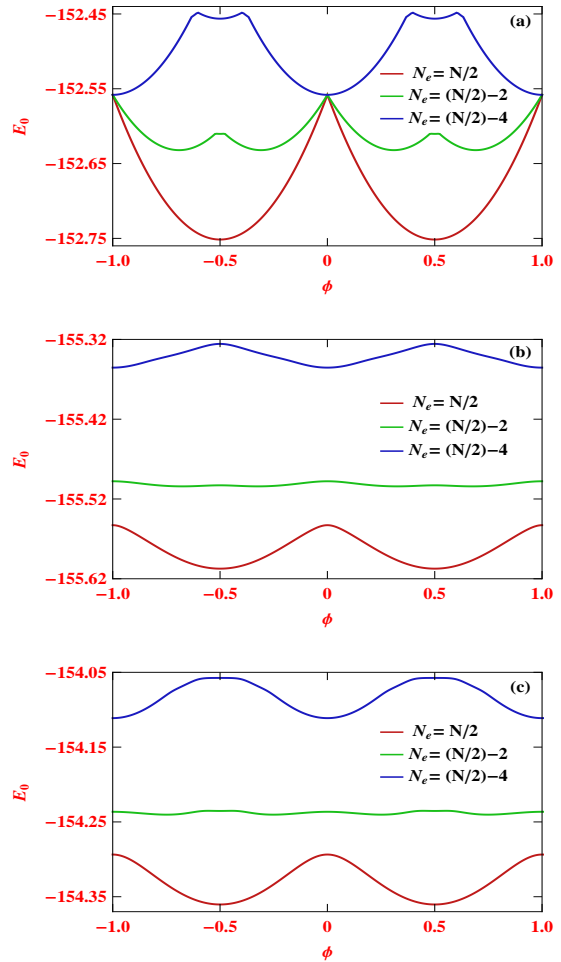}}\par}
\caption{(Color online). Ground state energy $E_0$ as a function of magnetic flux 
$\phi$ for three different filling factors where (a), (b) and (c) correspond 
to the identical meaning as given in Fig.~\ref{fig2}. Here we choose $N_s = 50$
and $N_r = 4$. For the disordered region we fix $W=1$.}
\label{fig3}
\end{figure}

Naturally, a question appears how ground state energy of such different kinds of
SWNTs depends on flux $\phi$ for different electron fillings, since the full 
spectrum shows many interesting patterns in these cases. To reveal this fact 
in Fig.~\ref{fig3} we present the dependence of ground state energy $E_0$ as a
function of magnetic flux $\phi$ for different band fillings for three 
different types of SWNTs considering $N_s=50$ and $N_r=4$. Several 
interesting features are observed associated with filling factor as well as 
spatial distribution of randomness i.e., whether the tube is partly or fully 
disordered. For the disordered free SWNT, ground state energy varies 
smoothly with $\phi$ exhibiting extrema at zero and integer multiples of 
$\phi_0$ in the half-filled band case 
(red curve of Fig.~\ref{fig3}(a)). The situation gets changed 
when the filling factor is less than half-filling, showing a sudden phase change 
around $\phi=\pm \phi_0/2$ (see green and blue lines of Fig.~\ref{fig3}(a)),
establishing a valley-like structure. Quite interestingly we see that the width
of this valley region gets increased with decreasing the filling factor. This is
the generic feature of such a multi-channel non-interacting system and not 
observed if we shrink it into a single-channel ring system. This valley-like 
behavior vanishes as long as impurities are introduced either in the full part 
or in sub-part of the SWNT (see the spectra given in Figs.~\ref{fig3}(b) 
and (c)). 
The slope of $E_0-\phi$ curve strongly depends on the filling factor for each 
SWNT, and comparing the curves shown in Figs.~\ref{fig3}(b) and (c) it 
is noticed that for the ODS SWNT change in slope of ground state energy 
with $\phi$ is quite higher than that of a fully disordered one. Here we present 
the results for one 
specific disorder strength (setting $W=1$), to set an example, and interesting 
behaviors are also observed for other disorder strengths. Those characteristics 
are directly reflected in circular current as discussed below.

\begin{figure}[ht]
{\centering \resizebox*{8cm}{13cm}{\includegraphics{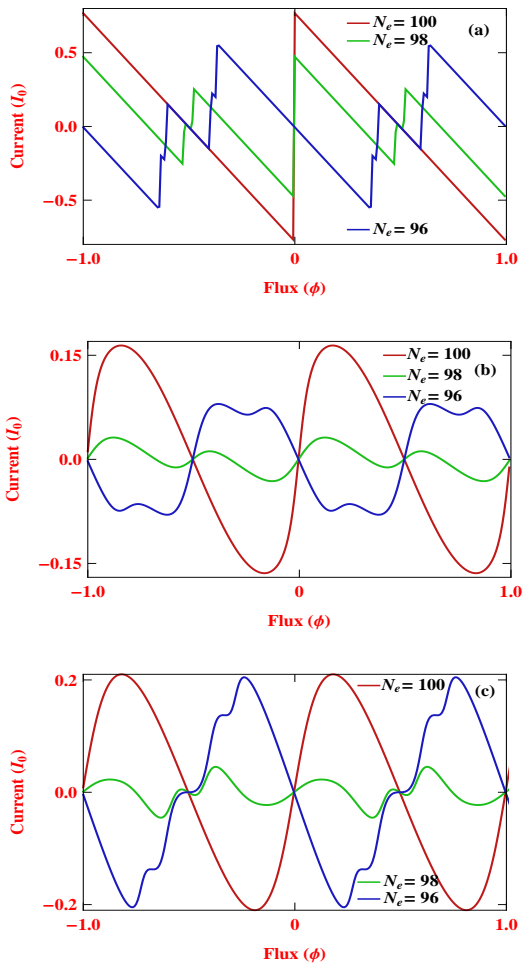}}\par}
\caption{(Color online). Current-flux characteristics for different filled
band cases considering the same set of parameter values as taken in 
Fig.~\ref{fig3}, where (a), (b) and (c) correspond to the ordered, 
fully disordered and ODS SWNTs, respectively.} 
\label{fig4}
\end{figure}

Figure~\ref{fig4} displays the variation of circular current as a function of
flux $\phi$ for the same set of parameter values as taken in Fig.~\ref{fig3}
to get a clear reflectance of energy variation on current. For the fully ordered
SWNT current shows saw-tooth like behavior with sharp transitions at integer
multiples of $\phi_0$ when the tube is half-filled. When the filling factor gets
reduced additional kink-like structures appear across $\pm \phi_0/2$ (green and
blue lines of Fig.~\ref{fig4}(a)) associated with the valleys in the $E_0$-$\phi$ 
curve, and the width of the kinks increases with more reduction of filling factor. 
Here it is important to note that this kink-like feature is no longer available 
in a single-channel non-interacting perfect AB ring for any filling factor. With
the inclusion of impurities all these kinks disappear (see Figs.~\ref{fig4}(b)
and (c)) following the $E_0$-$\phi$ curves. Both the spectra, Figs.~\ref{fig4}(b)
and (c), exhibit strong dependence of current on $N_e$ which suggests that wide
variation of current amplitude can be expected by regulating the filling factor.
This is indeed more transparent from the spectrum given in Fig.~\ref{fig4}(c),
and thus, ODS SWNT will be a very good candidate for regulating current 
amplitude, or in other words, electrical conductivity that may lead to the 
useful information in designing switching devices at nano-scale level. The 
other observation is that for the ODS SWNT current amplitude is 
comparatively higher than that of the fully disordered one for a specific 
filling factor. This seems to be expected as for a completely disordered 
nanotube more scattering takes place rather than the ODS one. 
It generates an interesting question that can we expect a different behavior
in current compared to the conventional spatially uniform disordered system if
we increase the disorder strength $W$. 

To answer this question let us have a look into the spectra given in 
\begin{figure}[ht]
{\centering \resizebox*{8cm}{8cm}{\includegraphics{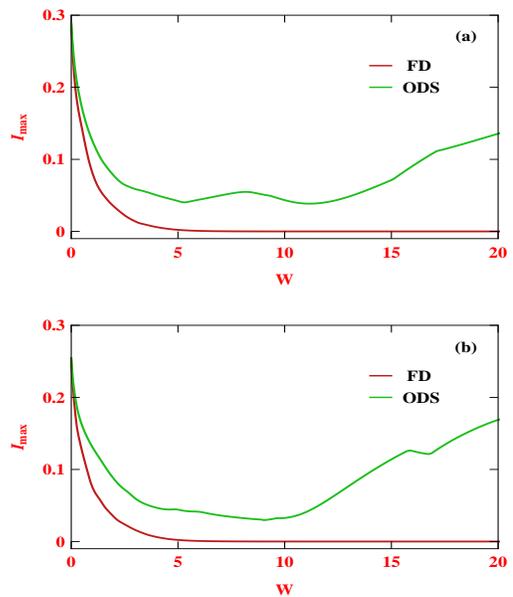}}\par}
\caption{(Color online). Maximum current amplitude ($I_{\mbox{\tiny max}}$) 
as a function of $W$ for ODS and fully disordered (FD) SWNTs considering
$N_s=50$ and $N_r=10$ where (a) and (b) correspond to $N_e=250$ and $248$, 
respectively. $I_{\mbox{\tiny max}}$ is calculated by taking the maximum 
absolute current within the flux range $0$ to $\phi_0$ for each $W$.}
\label{fig5}
\end{figure}
\begin{figure}[ht]
{\centering \resizebox*{8cm}{8cm}{\includegraphics{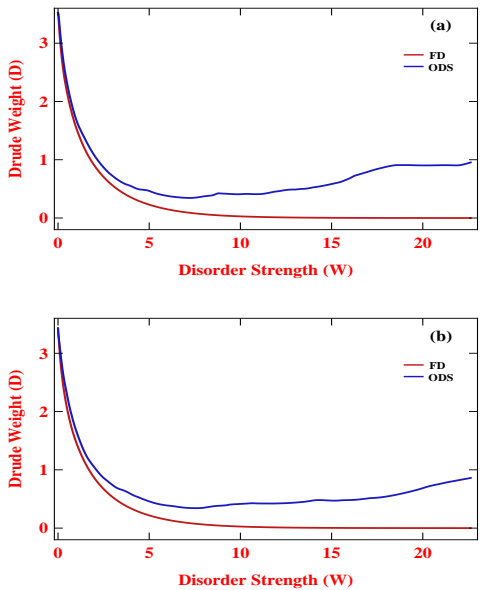}}\par}
\caption{(Color online). $D$-$W$ characteristics for ODS and FD SWNTs, where 
(a) $N_e = 250$ and (b) $N_e = 246$. $D$ is measured in unit of ($N/4\pi^2$).}
\label{fig6}
\end{figure}
Fig.~\ref{fig5} where the dependence of maximum current amplitude
($I_{\mbox{\tiny max}}$) is shown as a function of $W$ for both ODS and fully 
disordered (FD) SWNTs at two distinct filling factors. $I_{\mbox{\tiny max}}$
corresponds to the maximum absolute current within the flux range zero to one
flux-quantum (i.e., one complete period). A completely different scenario is
observed between the spatially uniform and non-uniform disordered SWNTs. For
the uniform disordered (viz, FD) case, current amplitude gradually decreases
with $W$ and eventually it drops to zero for large disorder strength. This is
quite natural since all the energy eigenstates are getting localized (Anderson
type localization) with increasing $W$. On the other hand, for the ODS SWNT 
initially we get decreasing nature of current
like what we get in a uniformly distributed disordered nanotube, but surprisingly
the current gets enhanced with the rise of disorder strength (green curves in
Fig.~\ref{fig5}). And in this strong disorder range current amplitude never 
decreases even in the limit of large $W$ rather it shows increasing tendency, 
which we confirm through our detailed and comprehensive numerical calculations. 
At extreme limit it will saturate. Thus we can classify two regions, weak and 
strong, depending on the strength of $W$ for the ODS SWNT where current, and 
hence, energy eigenstates behave completely differently. In the weak disorder
regime usual Anderson type localization is obtained, while for the other regime
\begin{figure}[ht]
{\centering \resizebox*{7.5cm}{9cm}{\includegraphics{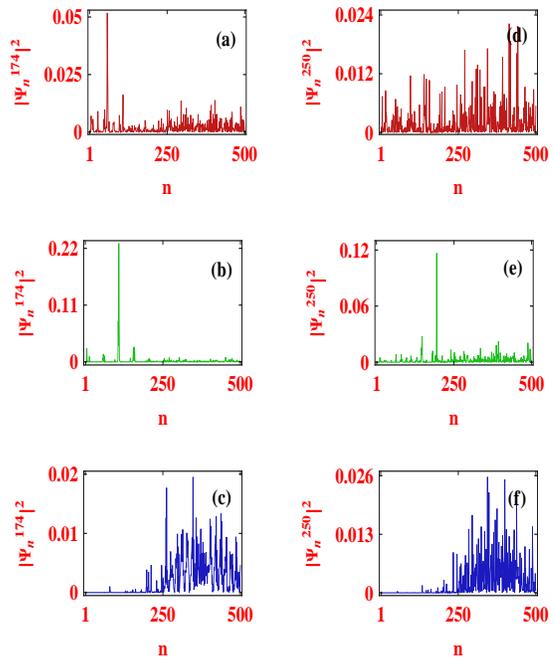}}\par}
\caption{(Color online). Probability distributions at different lattice sites
$n$ for an ODS SWNT for two eigenfunctions, $\Psi^{174}$ and $\Psi^{250}$, 
where the first, second and third rows correspond to $W=5$, $10$ and $20$, 
respectively. Here we set $N_s=50$ and $N_r=10$.}
\label{fig7}
\end{figure}
anti-localizing behavior is expected with increasing $W$. This phenomenon 
(anti-localization) can be analyzed from our forthcoming analysis. Before that
we want to check how this atypical nature of electronic localization plays the
role in electrical conductivity, and we discuss it by studying Drude weight
$D$. The results are given in Fig.~\ref{fig6} where two different filled band
cases are taken into account. Exactly similar pictures are obtained as described
in Fig.~\ref{fig5} which clearly emphasizes that the electrical conductivity 
can be tuned by regulating the strength of disorder in a spatially non-uniform
disordered material. That may give some important inputs towards device technology.

To explain physically the observed atypical nature let us focus on the  
spectra given in Fig.~\ref{fig7}, where we present the probability amplitudes
at different lattice sites ($n$) of an ODS SWNT choosing arbitrarily two 
distinct
eigenstates, $\psi^{174}$ and $\psi^{250}$, for three typical disorder strengths,
$W=5$, $10$ and $20$. These three values of $W$ are associated with the weak,
moderate and strong disorder regimes, respectively. Several interesting features 
are noticed from the spectra. In the regime of weak disorder, the probability 
of finding an electron at each lattice site of the $500$-site ODS SWNT is 
\begin{figure}[ht]
{\centering \resizebox*{8.5cm}{13.5cm}{\includegraphics{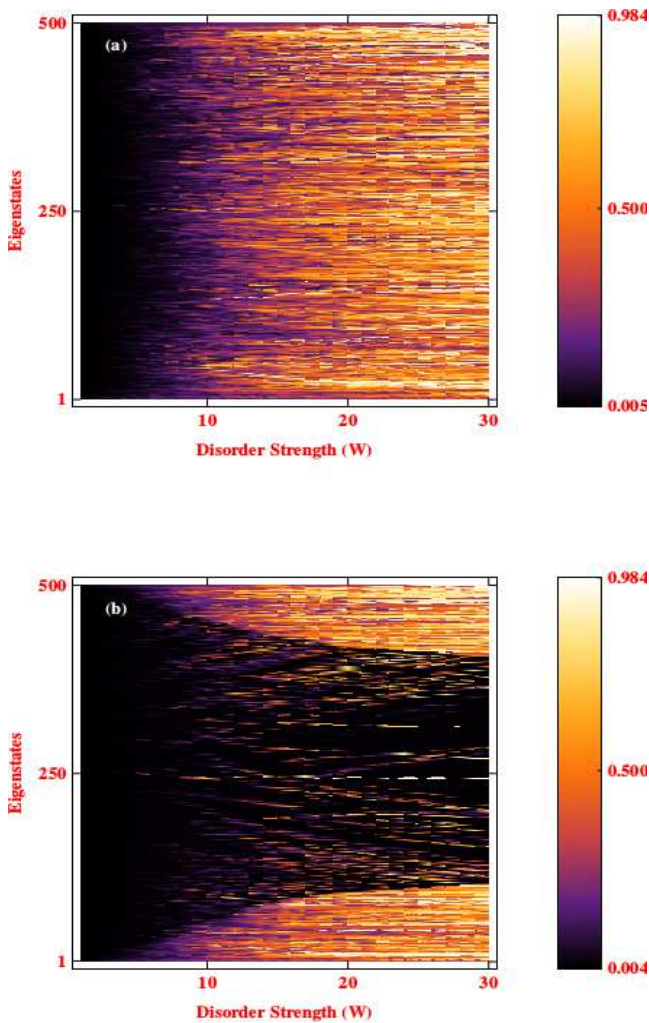}}\par}
\caption{(Color online). Inverse participation ratios (IPRs) of different energy
eigenstates as a function of $W$ where (a) and (b) correspond to FD and ODS 
SWNTs, respectively.}
\label{fig8}
\end{figure}
finite for both the two eigenstates, and it is also true for all other 
eigenstates, which yields that all the states are extended in nature. 
Definitely for a perfect SWNT 
we get more amplitudes at all atomic sites. Now for the typical $W$ ($W=10$)
where both the current and conductivity are too low (see the blue curves of 
Figs.~\ref{fig5} and \ref{fig6}), the probability amplitudes drop very close to 
zero almost for all the atomic sites apart from a single one where it gets a 
very high value. This is the common signature of a localized state as electron 
gets pinned at a particular site and does not able to hop along the system. 
Thus for this $W$ very small (not exactly zero, since absolute localization does
not take place due to finite size effect) current, and hence, electrical 
conductivity is obtained. This is the generic feature of Anderson 
localization~\cite{ander}. 
The peculiar behavior is obtained when the disorder strength gets increased.
For large $W$ we clearly see that in the disordered region (lower half of the 
nanotube i.e., from the sites $1$ to $250$ where impurities are introduced)
probability amplitudes are vanishingly small, whereas large amplitudes are 
noticed at all sites of the other half of the tube which is impurity free. It 
indicates that electrons can easily move in this ordered portion of the tube 
which essentially contributes to finite current. We also observe that the 
probability amplitudes in these impurity-free sites gradually increase with 
enhancing the strength $W$ of the impurity region. The whole process can be 
summarized as follows. The ODS SWNT is basically a coupled system where the 
disordered sector is directly coupled to the ordered one. As disorder favors 
scattering electrons start to get localized with increasing $W$ following 
Anderson localization~\cite{ander}. Thus more disorder causes more electronic 
localization
and therefore current gradually decreases and reaches to a minimum. So up to 
typical $W$ scattering effect dominates, but beyond this the coupling between 
the ordered and disordered regions gradually weakens which results suppressing 
the effect of disorder. Eventually for large disorder these two portions are
almost decoupled from each other, as clearly reflected from the probability 
amplitude spectra given in Fig.~\ref{fig7}. The phenomenon of decoupling can 
also be implemented following the analysis given in Ref.~\cite{div5}. Thus, 
when the ordered region is fully detached from the disordered one, electrons 
move freely in this region without getting scattered resulting a large current, 
and hence electrical conductivity. 

This decoupling effect in the limit of strong disorder enforces the appearance 
of LTD transition in such ODS SWNT. To implement it we compute inverse 
participation ratios (IPRs) of different energy eigenstates as a function of
$W$. The results are shown in Fig.~\ref{fig8} considering a $500$-site ODS 
nanotube (Fig.~\ref{fig8}(b)) along with the fully disordered one
(Fig.~\ref{fig8}(a)), for a better comparison between these two different kinds 
of nanotubes. For the FD nanotube conducting states are available only when the 
disorder strength is weak, while all these states get localized with increasing
$W$. The critical $W$ where all the states start to localize decreases with 
increasing the system size, which justifies the appearance of LTD transition at
weak disorder strength satisfying the Anderson prediction in 3D uncorrelated 
disordered model. Surprising behavior appears for the case of non-uniformly
distributed disordered system i.e., ODS SWNT. Even for large $W$ we always 
find the co-existence of localized and extended states. Thus we strongly 
argue that LTD transition persists in such system irrespective of disorder
strength which truly makes the ODS nanotube a special one compared to the FD 
nanotube.

\section{Closing Remarks}

To conclude, in the present communication we have shown that a spatially 
non-uniform disordered material exhibits a completely different behavior of
electronic localization compared to a uniformly distributed traditional 
disordered one. To justify this fact we have made a detailed analysis
of magneto-transport properties of non-interacting electrons considering 
an ODS SWNT along with a fully disordered SWNT. We have essentially 
characterized the full energy spectra, ground state energy at different band
fillings, filling dependent persistent currents and electrical conductivity.
Surprisingly we have seen that the ODS SWNT exhibits increasing circular
current, and hence, electrical conductivity with the enhancement of impurity
strength in the strong disorder regime. This is absolutely in contrast with
the spatially uniform traditional disordered material where always decreasing
nature with disorder is observed. This anomalous behavior has been clearly 
validated by inspecting the probability amplitudes at different lattice sites
of the ODS SWNT and found that with increasing impurity strength the ordered
sector gradually separated from the disordered one which results in suppression
of scattering effects and yields higher electrical conductivity. It suggests 
an opportunity of controlling electron mobility by tuning the doping strength
in doped materials. Finally, we have shown that {\em unlike 3D uncorrelated uniform
disordered material, the ODS SWNT exhibits mobility edge phenomenon and it 
persists even at strong disorder.} This phenomenon has been clearly justified by
investigating inverse participation ratios of different energy eigenstates.
Before we end, we would like to state that the present analysis may be useful
to analyze conducting properties of several doped materials.
 
\section{Acknowledgments}

MB and BM would like to thank Physics and Applied Mathematics Unit, Indian 
Statistical Institute, Kolkata, India for providing some facilities to work.

\end{document}